\documentclass[useAMS,usenatbib]{mnras}
\usepackage{multirow}
\usepackage{graphicx,color}
\usepackage{lscape}
\usepackage{pdflscape}
\newcommand{\etal }{{et al.} }
\newcommand{\msun}{\thinspace M_\odot}

\def\lesssim{\mathrel{\hbox{\rlap{\hbox{\lower4pt\hbox{$\sim$}}}\hbox{$<$}}}}
\def\gtrsim{\mathrel{\hbox{\rlap{\hbox{\lower4pt\hbox{$\sim$}}}\hbox{$>$}}}}
\newcommand{\cm}{\,{\rm cm}^{-3} }

\newcommand{\dfrac}[2]{{\displaystyle \frac{#1}{#2}} }

\newcommand{\zsun}{\thinspace Z_\odot}

\title[Outflow driving conditions]{Driving Conditions of Protostellar Outflows in Different Star-Forming Environments}

\author[K. ~Higuchi, \etal]
{Koki Higuchi$^{1}$\thanks{E-mail: higuchi.koki.054@s.kyushu-u.ac.jp (KH)}, Masahiro N. Machida$^{1}$ and Hajime Susa$^{2}$\\
$^{1}$Department of Earth and Planetary Sciences, Faculty of Sciences, Kyushu University, Fukuoka 812-8581, Japan\\
$^{2}$Department of Physics, Konan University, Okamoto, Kobe, 658-8501, Japan
}

\date{Accepted XXX. Received YYY; in original form ZZZ}
\pubyear{2019}

\begin{document}
\label{firstpage}
\pagerange{\pageref{firstpage}--\pageref{lastpage}}
\maketitle

\begin{abstract}
The evolution of collapsing clouds embedded in different star-forming environments is investigated using three-dimensional non-ideal magnetohydrodynamics simulations considering different cloud metallicities ($Z/\zsun$ = 0, 10$^{-5}$, 10$^{-4}$, 10$^{-3}$, 10$^{-2}$,
 10$^{-1}$ and 1) and ionisation strengths ($C_\zeta$=0, 0.01, 1 and 10, where $C_\zeta$ is a coefficient controlling the ionisation intensity and $C_\zeta=1$ corresponds to the ionisation strength of nearby star-forming regions). 
With all combinations of these considered values of $Z/\zsun$ and $C_\zeta$, 28 different star-forming environments are prepared and simulated.
The cloud evolution in each environment is calculated until the central density reaches $n\approx10^{16}\cm$ just before protostar formation, and the outflow driving conditions are derived.
An outflow appears when the (first) adiabatic core forms in a magnetically active region where the magnetic field is well coupled with the neutral gas. 
In cases where outflows are driven, their momentum fluxes are always comparable to the observations of nearby star-forming regions.  
Thus, these outflows should control the mass growth of the protostars as in the local universe. 
Roughly, an outflow appears when $Z/\zsun>10^{-4}$ and $C_\zeta \ge 0.01$.
It is expected that the transition of the star formation mode from massive stars to normal solar-type stars occurs when the cloud metallicity is enhanced to the range of $Z/\zsun\approx 10^{-4}$--$10^{-3}$, above which relatively low-mass stars would preferentially appear as a result of strong mass ejection.

\end{abstract}
\begin{keywords}
magnetohydrodynamics (MHD) -- stars: magnetic field -- stars: Population II -- stars: Population III
\end{keywords}

%%%%%%%%%%%%%%%%% BODY OF PAPER %%%%%%%%%%%%%%%%%%
\section{Introduction}
\label{sec:intro}
Protostellar outflows have been ubiquitously observed in nearby star-forming regions \citep[e.g.][]{Zhang05,Plunkett13}. 
Observations have confirmed a wide range of protostellar masses as the outflow driving sources.
In addition to low-mass stars, both extremely low-mass (i.e. proto-brown dwarfs; \citealt{Whelan18}) and very massive protostars \citep[e.g.][]{Beuther02,Maud15} also show protostellar outflows. 
Observations have also revealed the bipolarity and highly collimated structure of the protostellar outflows.
The outflow driving mechanism has been discussed since the 1980s and was an on-going mystery in the star formation process \citep[e.g.][]{pudritz83,pudritz86,uchida85,ouyed97}.
Very recent ALMA observations in which the outflow rotation and its driving region are spatially resolved are helping to unveil the outflow driving mechanism \citep{Bjerkeli16,Hirota17,Tabone17,Alves17}. 
In these observations, the authors concluded that the outflow can be explained by the so-called magnetohydrodynamic (MHD) disc wind model, in which the gas fluid flows outward along inclined magnetic field lines from the rotating disc \citep{blandford82}. 

Both observational and theoretical studies have predicted that protostellar outflows have a significant impact on present-day star formation \citep{McKee07}. 
Because such outflows sweep a large fraction of infalling gas and eject it into the interstellar space, they control the star formation efficiency and determine the final stellar mass \citep{Matzner00}.
In addition, the outflow transfers the excess angular momentum of the rotating disc and suppresses vigorous fragmentation \citep{tomisaka02,machida05,machida07,machida08}.
Thus, present-day star formation cannot be understood without considering outflows or MHD disc wind. 

Stars form in various galaxies and environments. 
Basically, except in very nearby star-forming regions in our galaxy, individual protostellar outflows and their driving sources cannot be resolved. 
Thus, it cannot be known whether protostellar outflows emerge in different star-forming environments (i.e. in other galaxies or during the early epochs of the universe). 
However, a protostellar outflow was recently observed in the Large Magellanic Cloud, which has a lower metallicity than our galaxy \citep{mcleod018}. 
Thus, outflows (or MHD disc winds) may appear around protostars formed in different star-forming environments (or different star formation processes).

The star formation process is divided into two phases: the gas collapsing and gas accretion phases. 
The former is the phase before protostar formation after star-forming cores begin to collapse, while the latter is the phase after protostar formation until the gas accretion onto the protostar stops. 
In primordial and present-day cases, the gas collapsing phase was firstly investigated by various authors (e.g. \citealt{Bromm02,Abel02,Yoshida08} for the primordial case; \citealt{tomisaka02,machida04,banerjee06} for the present-day case). 
Without understanding the gas collapsing phase, it is difficult to  understand the gas accretion phase following the gas collapsing phase because phenomena and structures appeared in the gas collapsing phase significantly affect the following gas accretion phase  \citep[e.g.][]{bate98,machida11,greif12}.
Thus, as a first step, we should focus on the gas collapsing phase even when investigating the star formation in various environments. 

The central density in collapsing clouds is usually used as an index of the cloud evolution in order to investigate the gas collapsing phase because the gas temperature (or pressure), attenuation of cosmic ray, ionization rate and magnetic resistivity, which focus on this study (see below),  strongly depend on the cloud (central) density. 
Note that the elapsed time after the cloud begins to collapse is not a better index describing the cloud evolution because the timescale of the high-density region is much shorter than that of the low-density region. 
Hereafter, we use the cloud (central) density to identify a specific  stage (or epoch) of the gas collapsing phase as done in past studies \citep[e.g.][]{omukai05,susa15}.

We get back to talking about the protostellar outflow.
The driving mechanism of outflow is closely related to the amplification and dissipation of the magnetic field because the Lorenz force plays a primary role in the MHD disc wind model \citep{blandford82,tomisaka98,tomisaka00,tomisaka02,banerjee06,machida08}. 
Because star-forming clouds are composed of weakly ionised plasma, the magnetic field dissipates by Ohmic dissipation and ambipolar diffusion and weakens in regions of high-density gas \citep{tomida15,tsukamoto15}. 
The diffusion rate of the magnetic field strongly depends on the ionisation degree, which is determined by both the chemical abundance of charged species and the strength of ionisation sources \citep[e.g.][]{okuzumi09,susa15}. 
For the present-day case, in a collapsing star-forming cloud, the magnetic field is coupled with the neutral gas in the density range of $n\lesssim10^{10}\cm$ or $n\gtrsim 10^{16}\cm$ and decoupled from the neutral gas in the range of $10^{10}\cm \lesssim n \lesssim 10^{16}\cm$ \citep{nakano02}.
The protostellar outflow can be driven by the magnetically active region of the rotating disc, in which the magnetic field is well coupled with the neutral gas.

As described above, the dissipation of the magnetic field must be considered to investigate the outflow driving mechanism, and the dissipation rate is determined by the chemical (or metal) abundance and ionisation intensity, which should differ in every star-forming environment. 
For example, the magnetic field is well coupled with the neutral gas everywhere in a primordial environment because dust grains, which absorb charged particles, do not exist in such a environment \citep{maki04,maki07}. 
On the other hand, the magnetically inactive region would dominate in the star-forming cloud when the ionisation intensity, such as that provided by cosmic rays, is very weak or does not exist \citep{susa15}.

Recently, \citet{tanaka18} noted that the MHD disc wind, if it emerges, is the primary feedback mechanism determining the final stellar mass, even in lower-metallicity environments or the early universe. 
They also claimed that the MHD disc wind dominates the radiation feedback as long as the protostellar mass is smaller than a few hundreds of solar masses \citep{kuiper18}.
Their results indicate that when the MHD disc wind appears, it significantly affects the formation of stars regardless of the environment. 
However, an outflow does not appear if the dissipation of the magnetic field, which was not considered by \citet{tanaka18}, is very effective in star-forming clouds.

In the present series of studies, we have previously investigated the dissipation and amplification of magnetic field in different star-forming environments \citep[][hereafter Paper I]{higuchi18}.
In Paper I, the evolution of the magnetic field in 36 different star-forming environments was investigated using three-dimensional non-ideal MHD simulations. 
However, that study focused only on spherically collapsing clouds, which were modelled as having initially very weak magnetic fields and very slow rotation rates to maintain their spherical structure. 
It should be noted that a strong magnetic field and/or rapid rotation changes the geometry of the collapsing cloud and the amplification rate of magnetic field. 
In addition, the formation of a rotationally supported disc and the emergence of an outflow complicate the magnetic field structure and make it difficult to investigate the amplification and dissipation of the magnetic field.

For these reasons, in Paper I, somewhat unrealistically weak magnetic fields and slow rotation rates were selected to only discuss the dissipation of the magnetic field or the effect of Ohmic dissipation and ambipolar diffusion in different star-forming environments. 
Following Paper I, this study investigates star formation in different environments using models of star-forming clouds with realistic magnetic field strengths and rotation rates.
Particular focus is placed on  the driving of outflows in the collapsing cloud embedded in different star-forming environments.
The outflow driving mechanism in collapsing clouds in present-day star formation has been investigated in many previous works \citep[e.g.][]{tomisaka02,machida08,price12,tomida13,bate14,tomida15,tsukamoto15,matsushita17}; however, the same mechanism in different environments has been investigated in only a few studies \citep{machida06,machida09}.  
In this study, outflow emergence is investigated in a wide parameter space of cloud metallicity and ionisation intensity,  which has not been performed in any previous study.

The remainder of the present paper is organised as follows.
The numerical settings are described in \S\ref{sec:methods}, and the results are presented in \S\ref{sec:results}. 
In addition to the Hall effect and magnetic braking catastrophe, the outflow driving conditions, angular momentum transfer, and transition of the star formation mode are discussed in \S\ref{sec:discussion}. 
Finally, a summary is presented in \S\ref{sec:summary}.

\section{Initial Settings and Parameters}
\label{sec:methods}
\begin{table*}
\centering
\caption{
Model parameters.
Columns 1 and 2 list the model number and model name, respectively.
Columns 3 through 5 list the ionisation strength $C_\zeta$, the metallicity $Z/\zsun$, and the mass-to-flux ratio $\mu_0$, respectively.
Columns 6 through 10 list the initial magnetic field strength $B_0$, the initial angular velocity $\Omega_0$, the cloud mass $M_{\rm cl}$, the isothermal temperature $T_{\rm cl}$, and the cloud radius $r_{\rm cl}$, respectively.
Column 11 lists whether outflow emerges ($\circ$, $\diamond$) or not ($\times$), where the symbol $\circ$ indicates the outflow is driven in the high-density region (high-density outflow; $n>10^{10}\cm$) and $\diamond$ means that the outflow is driven in the low-density region (low-density outflow; $n<10^{10}\cm$).
}
\begin{tabular}{|c||c|c|c|c|c|c|c|c|c|c|c|c|} \hline
      &  Model & $C_{\zeta}$ & $Z/\zsun$ & $\mu_0$ & $B_0$ $[\rm{\mu G}]$ & $\Omega_0$ $[\rm{s^{-1}}]$ & $M_{\rm{cl}}$ $[M_{\odot}]$ & $T_{\rm{cl}}$ $[\rm{K}]$ & $r_{\rm{cl}}$ $[\rm{AU}]$ & Outflow \\ \hline

%%%%%%%%%%%%%%%%%%%%%%%%%%%%%%%%%%%%%%%%%%%%%%%%%%%%%%%%%%%%%%%%%%%%%%%%%%%%%
%                             C_\zeta = 0
%%%%%%%%%%%%%%%%%%%%%%%%%%%%%%%%%%%%%%%%%%%%%%%%%%%%%%%%%%%%%%%%%%%%%%%%%%%%%
 1&      I0ZP &     & $0$       &   & 34.1 & $1.31 \times 10^{-14}$ & $1.08 \times 10^{4}$ & 198 & $4.91 \times 10^{5}$ & $\times$ &\\

 2&      I0Z5 &     & $10^{-5}$ &   & 33.8 & $1.31 \times 10^{-14}$ & $1.05 \times 10^{4}$ & 194 & $4.87 \times 10^{5}$ & $\times$ &\\

 3&      I0Z4 &     & $10^{-4}$ &   & 31.9 & $1.31 \times 10^{-14}$ & $8.75 \times 10^{3}$ & 172 & $4.59 \times 10^{5}$ & $\times$ &\\

 4&      I0Z3 & $0$ & $10^{-3}$ & 3 & 24.6 & $1.31 \times 10^{-14}$ & $3.98 \times 10^{3}$ & 103 & $3.52 \times 10^{5}$ & $\times$ &\\

 5&      I0Z2 &     & $10^{-2}$ &   & 9.83 & $1.35 \times 10^{-14}$ & $2.27 \times 10^{2}$ & 16.4 & $1.33 \times 10^{5}$ & $\times$ &\\

 6&      I0Z1 &     & $10^{-1}$ &   & 10.3 & $1.62 \times 10^{-14}$ & $1.26 \times 10^{2}$ & 18.1 & $9.67 \times 10^{4}$ & $\circ$ &\\

 7&      I0Z0 &     & $1$       &   & 5.76 & $1.78 \times 10^{-14}$ & $15.2$               & 5.65 & $4.49 \times 10^{4}$ & $\circ$ &\\ \hline
%%%%%%%%%%%%%%%%%%%%%%%%%%%%%%%%%%%%%%%%%%%%%%%%%%%%%%%%%%%%%%%%%%%%%%%%%%%%%
%                             C_\zeta = 0.01
%%%%%%%%%%%%%%%%%%%%%%%%%%%%%%%%%%%%%%%%%%%%%%%%%%%%%%%%%%%%%%%%%%%%%%%%%%%%%
 8&      I001ZP &        & $0$       &   & 28.4 & $1.31 \times 10^{-14}$ & $6.20 \times 10^{3}$ & 140 & $4.09 \times 10^{5}$ & $\times$ &\\

 9&      I001Z5 &        & $10^{-5}$ &   & 25.1 & $1.31 \times 10^{-14}$ & $6.03 \times 10^{3}$ & 136 & $4.05 \times 10^{5}$ & $\times$ &\\

 10&     I001Z4 &        & $10^{-4}$ &   & 26.2 & $1.31 \times 10^{-14}$ & $4.88 \times 10^{3}$ & 117 & $3.77 \times 10^{5}$ & $\times$ &\\

 11&     I001Z3 & $0.01$ & $10^{-3}$ & 3 & 20.0 & $1.31 \times 10^{-14}$ & $2.15 \times 10^{3}$ & 68.0 & $2.87 \times 10^{5}$ & $\circ$ &\\

 12&     I001Z2 &        & $10^{-2}$ &   & 9.85 & $1.35 \times 10^{-14}$ & $2.30 \times 10^{2}$ & 16.5 & $1.34 \times 10^{5}$ & $\circ$ &\\

 13&     I001Z1 &        & $10^{-1}$ &   & 10.4 & $1.62 \times 10^{-14}$ & $1.28 \times 10^{2}$ & 18.2 & $9.72 \times 10^{4}$ & $\circ$ &\\

 14&     I001Z0 &        & $1$       &   & 5.76 & $1.78 \times 10^{-14}$ & 15.2                 & 5.64 & $4.49 \times 10^{4}$ & $\circ$ &\\ \hline
%%%%%%%%%%%%%%%%%%%%%%%%%%%%%%%%%%%%%%%%%%%%%%%%%%%%%%%%%%%%%%%%%%%%%%%%%%%%%
%                             C_\zeta = 1
%%%%%%%%%%%%%%%%%%%%%%%%%%%%%%%%%%%%%%%%%%%%%%%%%%%%%%%%%%%%%%%%%%%%%%%%%%%%%
 15&     I1ZP &     & $0$       &   & 12.1 & $1.31 \times 10^{-14}$ & $4.79 \times 10^2$ & 24.9 & $1.74 \times 10^{5}$ & $\diamond$ &\\

 16&     I1Z5 &     & $10^{-5}$ &   & 12.1 & $1.31 \times 10^{-14}$ & $4.82 \times 10^2$ & 25.1 & $1.74 \times 10^{5}$ & $\diamond$ &\\

 17&     I1Z4 &     & $10^{-4}$ &   & 12.4 & $1.31 \times 10^{-14}$ & $5.09 \times 10^2$ & 26.0 & $1.77 \times 10^{5}$ & $\times$ &\\

 18&     I1Z3 & $1$ & $10^{-3}$ & 3 & 12.7 & $1.31 \times 10^{-14}$ & $5.43 \times 10^2$ & 27.3 & $1.81 \times 10^{5}$ & $\circ$ &\\

 19&     I1Z2 &     & $10^{-2}$ &   & 12.1 & $1.34 \times 10^{-14}$ & $4.39 \times 10^2$ & 25.0 & $1.66 \times 10^{5}$ & $\circ$ &\\

 20&     I1Z1 &     & $10^{-1}$ &   & 10.9 & $1.59 \times 10^{-14}$ & $1.58 \times 10^2$ & 20.1 & $1.06 \times 10^{5}$ & $\circ$ &\\

 21&     I1Z0 &     & $1$       &   & 6.11 & $1.78 \times 10^{-14}$ & 18.0               & 6.34 & $4.75 \times 10^{4}$ & $\circ$ &\\ \hline
%%%%%%%%%%%%%%%%%%%%%%%%%%%%%%%%%%%%%%%%%%%%%%%%%%%%%%%%%%%%%%%%%%%%%%%%%%%%%
%                             C_\zeta = 10
%%%%%%%%%%%%%%%%%%%%%%%%%%%%%%%%%%%%%%%%%%%%%%%%%%%%%%%%%%%%%%%%%%%%%%%%%%%%%
 22&     I10ZP &      & $0$       &   & 13.5 & $1.31 \times 10^{-14}$ & $6.56 \times 10^2$ & 31.0 & $1.93 \times 10^{5}$ & $\diamond$ &\\

 23&     I10Z5 &      & $10^{-5}$ &   & 13.6 & $1.31 \times 10^{-14}$ & $6.64 \times 10^2$ & 31.2 & $1.94 \times 10^{5}$ & $\diamond$ &\\

 24&     I10Z4 &      & $10^{-4}$ &   & 14.0 & $1.32 \times 10^{-14}$ & $7.25 \times 10^2$ & 33.1 & $1.99 \times 10^{5}$ & $\times$ &\\

 25&     I10Z3 & $10$ & $10^{-3}$ & 3 & 15.3 & $1.32 \times 10^{-14}$ & $9.39 \times 10^2$ & 39.6 & $2.17 \times 10^{5}$ & $\circ$ &\\

 26&     I10Z2 &      & $10^{-2}$ &   & 15.3 & $1.34 \times 10^{-14}$ & $8.67 \times 10^2$ & 39.6 & $2.09 \times 10^{5}$ & $\circ$ &\\

 27&     I10Z1 &      & $10^{-1}$ &   & 12.6 & $1.55 \times 10^{-14}$ & $2.74 \times 10^2$ & 26.8 & $1.29 \times 10^{5}$ & $\circ$ &\\

 28&     I10Z0 &      & $1$       &   & 8.03 & $1.78 \times 10^{-14}$ & 40.1               & 11.0 & $6.24 \times 10^{4}$ & $\circ$ &\\ \hline

    \end{tabular}
%}
\label{table:1}
\end{table*}
Because the initial settings and parameters are almost the same as in Paper I, they are only briefly described in this section. 
In the same manner as in Paper I, star-forming clouds are prepared in different environments with different combinations of two parameters: the cloud metallicity $Z/\zsun$ and ionisation strength $C_\zeta$ (see Eqs.~(1)--(4) of Paper I). 
The model names and the values of $C_\zeta$ and $Z/\zsun$ are listed in Table~\ref{table:1}. 
Seven different cloud metallicities of $Z/\zsun=0$, 10$^{-5}$, 10$^{-4}$, 10$^{-3}$, 10$^{-2}$, 10$^{-1}$, and 1 and four different ionisation strengths of $C_\zeta=0$, 0.01, 1 and 10 are adopted in these models.  
All possible pairs of these parameter values yield a total of  28 models. 
The clouds with $Z/\zsun=0$ are prepared to investigate star formation in a primordial environment (or in the early universe), whereas the clouds with $Z=\zsun$ correspond to the conditions of present-day star formation. 
Because the metallicity increases with time, $Z/\zsun$ is used as an index of time (or the age) of the universe. The ionisation intensity is crucial in the determination of the ionisation degree, which controls the dissipation  of the magnetic field \citep{susa15}. 
The models with $C_\zeta=1$ have a cosmic ray rate of $\zeta_{\rm CR}=10^{-17}$\,s$^{-1}$ (for details, see Eqs.~(1)--(4) of Paper I). 
Note that because the cosmic ray attenuation is considered, $\zeta_{\rm CR}$ gradually decreases as the cloud density increases. 
In addition to cosmic rays, short- and long-lived radioactive elements are also considered as ionisation sources (see Paper I).   
There is no ionisation intensity for models with $C_\zeta=0$, corresponding to a primordial environment with no ionisation sources around the star-forming clouds, which were not polluted with metals by past supernovae.
On the other hand, models with $C_\zeta=10$ are prepared to consider vigorous star-forming environments, such as star-burst galaxies, where cosmic rays are expected to be abundantly supplied by frequent supernovae \citep{lacki14}.

As the initial state of star-forming clouds, a critical Bonnor--Ebert (B.E.) density profile \citep{ebert55,bonnor56} is adopted for each model. 
Note that the B.E. density profile or B.E. sphere is usually used as the initial condition of star-forming clouds \citep[e.g.][]{matsumoto04, machida06,machida06b, banerjee06}.
The B.E. density profile is determined by the central density $n_{\rm c,0}$ and isothermal temperature $T_{\rm cl}$. 
The initial central density is set to $n_{\rm c,0}=10^4\cm$ for all models.
The temperature $T_{\rm cl}$ of each cloud is determined as the result of a one-zone calculation \citep[for details, see][and Paper I]{susa15}, and the results $T_{\rm cl}$ are given in Table~\ref{table:1}.
The cloud radius $r_{\rm cl}$, which depends on the initial cloud temperature, is also given in Table~\ref{table:1}.
To promote cloud contraction, the density is set to 1.8 times to the critical B.E. density profile \citep{machida13}.
The initial cloud mass for each model is also listed in Table~\ref{table:1}.
Although the initial clouds have different radii and masses with different metallicities, the ratio $\alpha_0$ of thermal to gravitational energy, which significantly affects the cloud collapse \citep[e.g.][]{miyama84,tsuribe99a,tsuribe99b}, is the same for all models ($\alpha_0=0.47$).
In addition, the ratio of rotational to gravitational energy in the initial cloud is set to $\beta_0=1.84 \times 10^{-2}$ for all models \citep{goodman93,caselli02}.
The initial magnetic field strength in each cloud is defined to satisfy $\mu_0 = 3$ \citep{troland08,crutcher10,ching17}. The parameter $\mu_0$ is the mass-to-flux ratio of the initial cloud normalised by the critical value and is defined as
\begin{equation}
\mu_0 = \frac{\left( M/\Phi \right)}{\left( M/\Phi \right)_{\rm cri}},
\end{equation}
where $M$ and $\Phi$ are the mass and magnetic flux of the initial cloud, respectively, and $(M/\Phi)_{\rm cri}$ is the ratio of the critical values of these parameters, which is $(M/\Phi)_{\rm cri} \equiv (2 \pi G^{1/2})^{-1}$ \citep{nakano78}.
The direction of the magnetic field vector is parallel to the rotation vector ($z$-axis) in the initial cloud, in which a uniform magnetic field and rigid rotation are imposed.

As described above, the initial settings in this study are almost the same as in Paper I. 
The parameters that differed are the magnetic field strength and rotation rate of the initial cloud. 
In Paper I, unrealistically weak magnetic fields and slow rotation rates were adopted to focus on the dissipation rate of the magnetic field and suppress the outflow driving, anisotropic collapse, disc formation, and fragmentation, as described in \S\ref{sec:intro}.
In contrast, this study focuses on the effect of the magnetic field and rotation in the collapsing cloud.
Thus, for the present-day models (i.e. models having $Z=\zsun$), the adopted magnetic field strength and rotation rate should be similar to those observed in (nearby) star-forming regions. 
However, the magnetic field strengths and rotation rates in the early universe or other environments are not known. 
Then, the ratios of both the magnetic and rotational energies to the gravitational energy are assumed to be the same for all models. 
Therefore, among the models with different metallicities $Z/\zsun$, although the magnetic field strengths and rotation rates differ, the non-dimensional parameters $\alpha$, $\beta$, and $\mu$ are the same.
We consider that this is the only way to fairly investigate the evolution of the collapsing cloud with different masses and radii. 
The magnetic strength and rotation rate for each model are also listed in Table~\ref{table:1}.

The cloud collapse for each model in Table~\ref{table:1} is calculated using three-dimensional non-ideal MHD nested grid simulations, in which Ohmic dissipation and ambipolar diffusion terms are implemented \citep{machida18}. 
Because the numerical method is the same as in Paper I, it is only briefly described here. 
In the simulations, the gas pressure and Ohmic and ambipolar resistivities are set based on the values in a pre-calculated table \citep{susa15}, in which chemical reactions (or ionisation degree, electronic conductivities, and resistivities), density, and temperature (or gas pressure) were  calculated self-consistently using the one-zone model with the magnetic field strength as a parameter (see also Paper I). 
In the nested grid code, each grid is composed of $(i,j,k) = (64, 64, 32)$, and mirror symmetry is applied with respect to the $z$-axis \citep[for a detailed description of the nested grid, see][]{machida04,machida05,machida07,machida08}. 
Finer grids are embedded in a coarser grid, and the finer grids are successively generated such that the resolution of Jeans length is at least 16 cells \citep{truelove97}. 
With this code, the cloud evolution is calculated until the central density reaches $n_{\rm c}=10^{16}\cm$. 
Note that, for some models, the calculation is stopped at $n<10^{16}\cm$ because the time step of the magnetic dissipation becomes extremely short.
As described in \S\ref{sec:intro}, we use the central number density as a index of the cloud evolution. 
In this study, we only calculate the gas collapsing phase during which the central density continues to increase.
Thus,  the central density can distinctly trace each stage of collapsing clouds. 

%%%%%%%%%%%%%%%%%%%%%%%%%%%%%%%%%%%%%%%%%%%%%%%%%%%%%%%%%%%%%%%%%%%%%%%%%%%%%
%                                  results
%%%%%%%%%%%%%%%%%%%%%%%%%%%%%%%%%%%%%%%%%%%%%%%%%%%%%%%%%%%%%%%%%%%%%%%%%%%%%
\section{Results}
\label{sec:results}
\subsection{Typical Models}
\label{sec:model}

%%%%%%%%%%
% Fig. 1
%%%%%%%%%%
\begin{figure*}
    \includegraphics[scale=0.75]{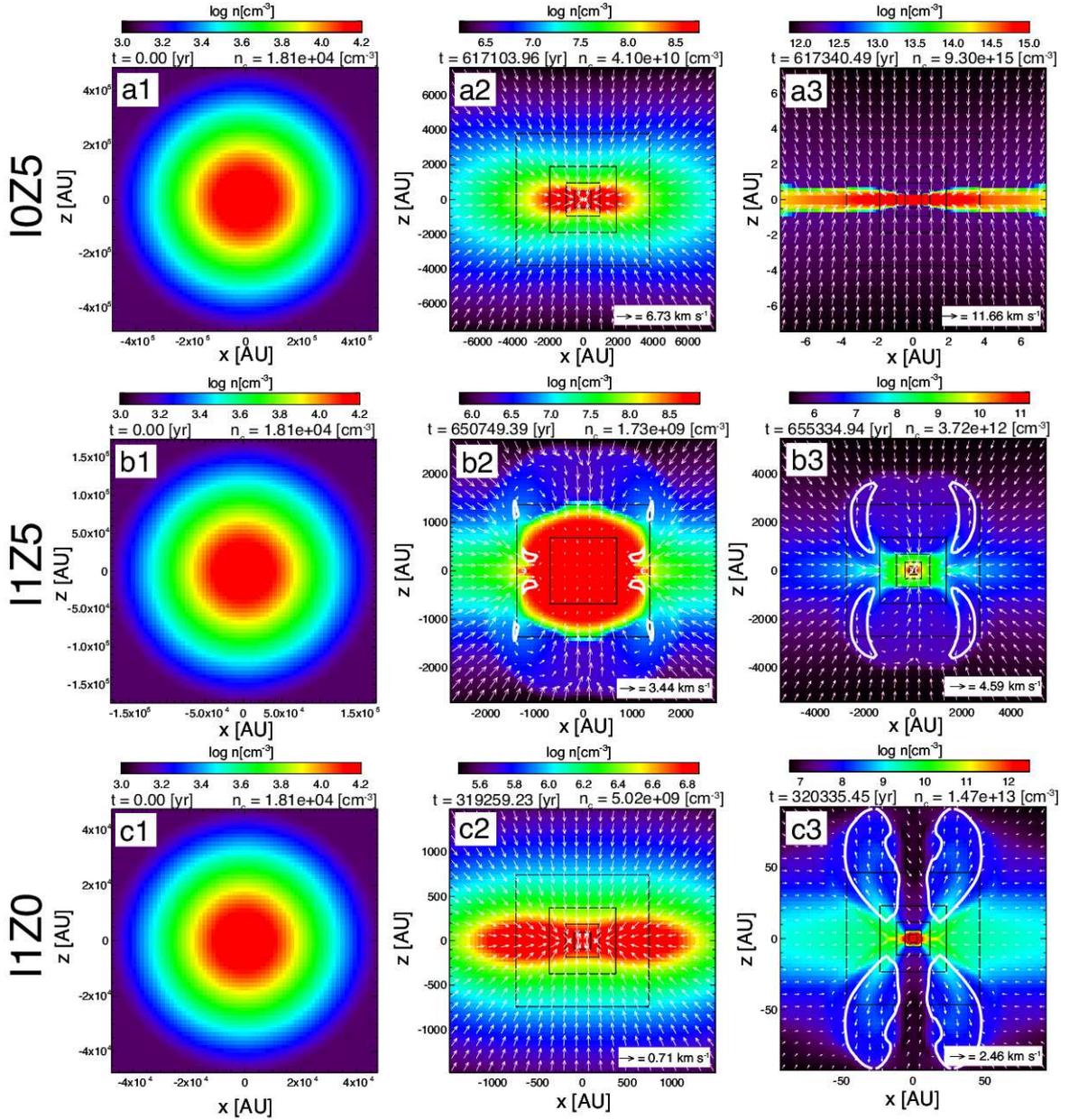}
\caption{
Time sequences of models I0Z5 (a1, a2, a3), I1Z5 (b1, b2, b3), and I1Z0 (c1, c2, c3).
The density (colour) and velocity (arrows) distributions in the $y=0$ plane are plotted in each panel.
The white solid line is the boundary between inflowing and outflowing regions. 
The elapsed time and central number density are given in above each panel.
The box scale differs in each panel.
The black squares in each panel indicate the grid boundaries.
}
    \label{fig:mu3_time}
\end{figure*}

First, the evolution and structures of three typical models are presented. 
Fig.~\ref{fig:mu3_time} shows the time sequences of models I0Z5, I1Z5 and I1Z0.
The figure indicates that the Lorentz and centrifugal forces produce a disc-like structure.
\footnote{
We call the flattened structure appeared in the collapsing cloud  `the disk-like structure,' which is different from the rotationally supported disk and Keplerian disk.
The disk-like structure continues to contact during the calculation. 
}
An outflow appeared in the collapsing cloud for models I1Z5 and I1Z0, whereas an outflow did not appear until the end of the calculation for model I0Z5.
Even among the models showing an outflow, the outflow driving regions and epochs differed considerably. 
For example, the outflow emerged at a lower density (or earlier epoch) for model I1Z5 than model I1Z0.
The difference in the outflow size between models I1Z5 and I1Z0 (Fig.~\ref{fig:mu3_time}{\it }) is just reflected  by the difference in the size of the outflow driving regions (or epochs  and densities, for details, see \S\ref{sec:outflow}).

\begin{figure*}
    \includegraphics[scale=0.75]{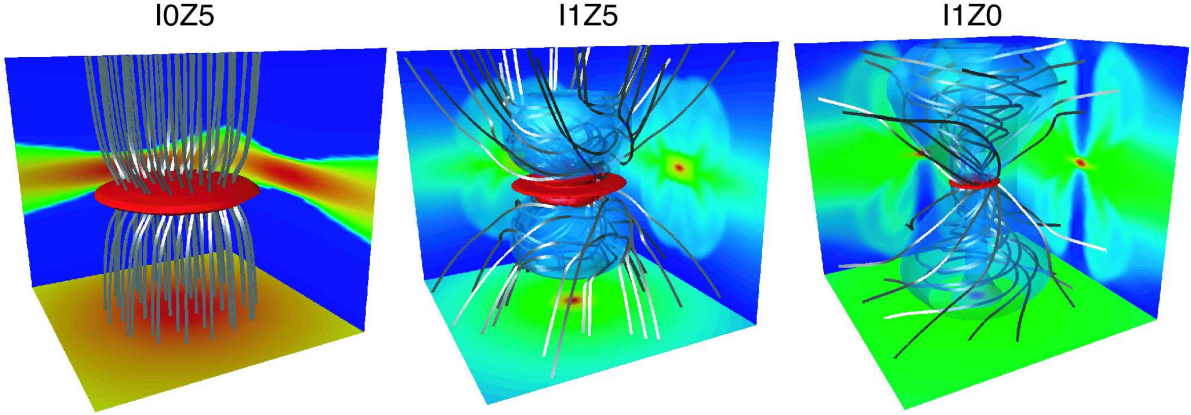}
\caption{
Three-dimensional view of models I0Z5 (left), I1Z5 (middle), and I1Z0 (right).
The spatial scale for each panel is the same as in the right panels of Fig.~\ref{fig:mu3_time}.
The magnetic field lines and high-density region are plotted as silver lines and a red contour, respectively. 
The outflow is represented by the blue surface, inside which the gas is outflowing from the centre. 
The density distribution on the $x=0$, $y=0$, and $z=0$ planes are projected on each wall surface. 
}
    \label{fig:3d}
\end{figure*}

Fig.~\ref{fig:3d} shows a three-dimensional view of the magnetic field lines, outflow, and density structure for the same models as in Fig.~\ref{fig:mu3_time}. 
A thin disc-like structure appeared in all three models, but outflows (blue surfaces) emerged only in models I1Z5 and I1Z0. 
The magnetic field lines were strongly twisted in models I1Z5 and I1Z0, whereas the field lines were almost unidirectionally aligned in model I0Z5. 
As described below, the magnetic field dissipates and is not well coupled with the neutral gas in model I0Z5. 
Thus, although the magnetic field lines are somewhat distorted, they have a simple configuration in the  model. 
In contrast, when the magnetic field is well coupled with the neutral gas, the magnetic field lines are strongly twisted, as in models I1Z5 and I1Z0.
In this case, the magnetic field is amplified, and the outflow is driven by the disc-like structure (middle and right panels of Fig.~\ref{fig:3d}).  
%%%%%%%%%%
% Fig. 2
%%%%%%%%%%

%%%%%%%%%%
% Fig. 3
%%%%%%%%%%
\subsection{Amplification of Magnetic Field for All Models}
\begin{figure*}
    \includegraphics[scale=0.6]{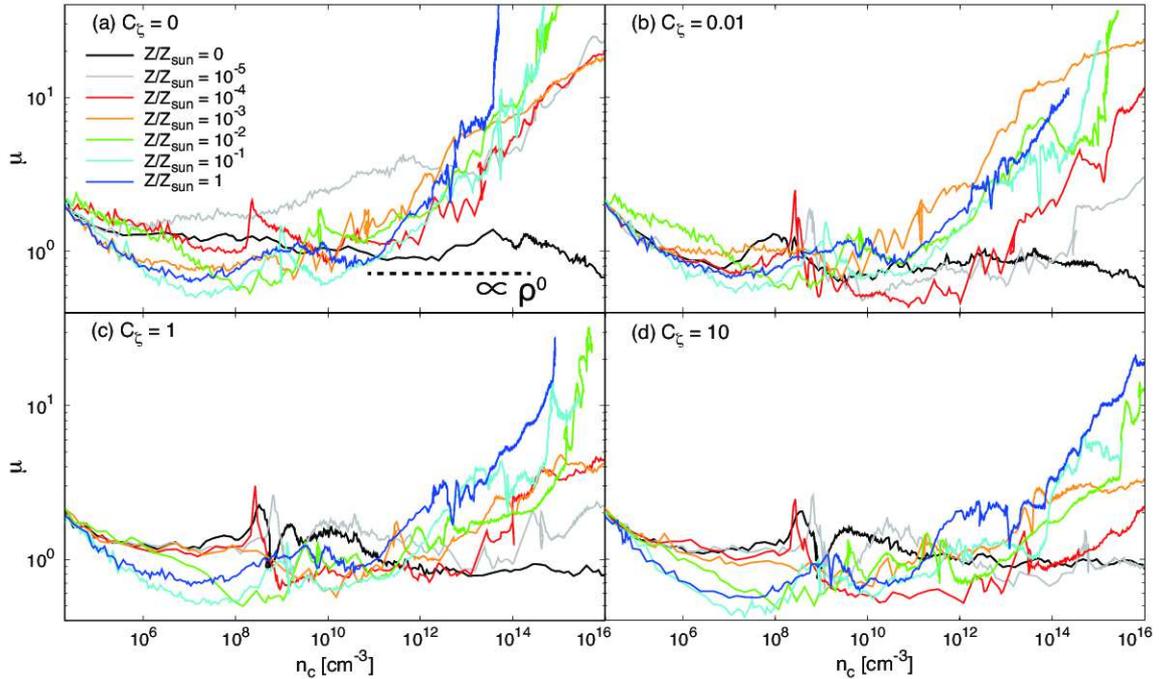}
\caption{
Mass-to-flux ratios at the centre of the collapsing cloud for models with different metallicities ($Z/\zsun= 0$--$1$) plotted against the central density for different ionisation strengths $C_\zeta$ of ({\it a}) $0$, ({\it b}) $0.01$, ({\it c}) $1$, and ({\it d}) $10$.
}
    \label{fig:mu3}
\end{figure*}
As described in Paper I, the mass-to-flux ratio $\mu$, which was estimated for $\rho>0.1\rho_{\rm c}$ ($\rho_{\rm c}$ is the central mass density of the collapsing cloud), is a useful index to compare magnetic field strengths in the collapsing clouds among different models. The mass-to-flux ratios for models with different metallicities and ionisation strengths are plotted against the central number density in Fig.~\ref{fig:mu3}.
This figure indicates that the amplification of the magnetic field strongly depends on the metallicity $Z/\zsun$ \citep[see also][Paper I]{susa15}.
The magnetic field strengths in lower-metallicity clouds tend to be less dissipated than those in higher-metallicity clouds. 
In addition, the amplification of the magnetic field also depends on the ionisation strength $C_\zeta$.

As shown in Fig.~\ref{fig:mu3}, in the zero-metallicity case ($Z/\zsun=0$; solid black line), the mass-to-flux ratio was almost constant (i.e. $\mu \propto \rho^0$).
The relationship between the magnetic field strength $B$ and the mass-to-flux ratio $\mu$ according to Paper I is explained here.  
The mass-to-flux ratio can be roughly described as $\mu \approx M /B L^2$, where the Jeans mass $M_{\rm J}$ and Jeans wavelength $R_{\rm J}$ are adopted as the typical mass $M$ and length $L$, respectively.
Assuming that $T \propto \rho^{\gamma -1}$, the Jeans mass and wavelength satisfy the following proportionalities: $ M_{\rm{J}} \propto \rho^{(3\gamma -4)/2} $ and $ R_{\rm{J}} \propto \rho^{(\gamma -2)/2}$.
Because the polytropic index $\gamma$ is almost 1 for $Z/\zsun = 0$ \citep{omukai98,omukai00,omukai05,omukai10}, the mass-to-flux ratio satisfies $\mu \propto \rho^{1/2} B^{-1}$.
Using the calculation result of $\mu \propto \rho^0$ for the primordial case, the dependence of the magnetic field on the density can be estimated as $B\propto \rho^{1/2}$. 
Furthermore, when the magnetic field is well coupled with the neutral gas and the cloud collapses while maintaining a disc-like structure, the magnetic field is amplified as $B\propto \rho^{1/2}$ \citep{scott80}. 
Thus, as long as $\mu \propto \rho^0$ holds, the magnetic field is well coupled with the neutral gas. 
Therefore, Fig.~\ref{fig:mu3} indicates that, for the primordial case, the magnetic field is well coupled with the neutral gas in all ranges, which is consistent with \citet{susa15} and Paper I.  
When the dissipation of the magnetic field becomes effective, the mass-to-flux ratio begins to increase.
Thus, Fig.~\ref{fig:mu3} also indicates that, except for models with $Z/\zsun = 0$,  the dissipation of the magnetic field becomes effective when the number density exceeds $10^{10}$--$10^{12}\cm$, at which the mass-to-flux ratios begin to increase.

Figs.~\ref{fig:mu3}{\it a}--{\it d} also show that overall, the mass-to-flux ratio gradually decreases as the ionisation strength $C_\zeta$ increases in all the models except in the case of $Z/\zsun = 0$. 
This indicates that the magnetic field is stronger in models with higher $C_\zeta$ than in models with lower $C_\zeta$. 
This is natural because as the ionisation intensity strengthens, the ionisation rate increases and the dissipation rate of the magnetic field decreases (see also Paper I). 

%%%%%%%%%%
% Fig. 4
%%%%%%%%%%

%\begin{landscape}
\begin{figure*}
    \includegraphics[scale=0.60]{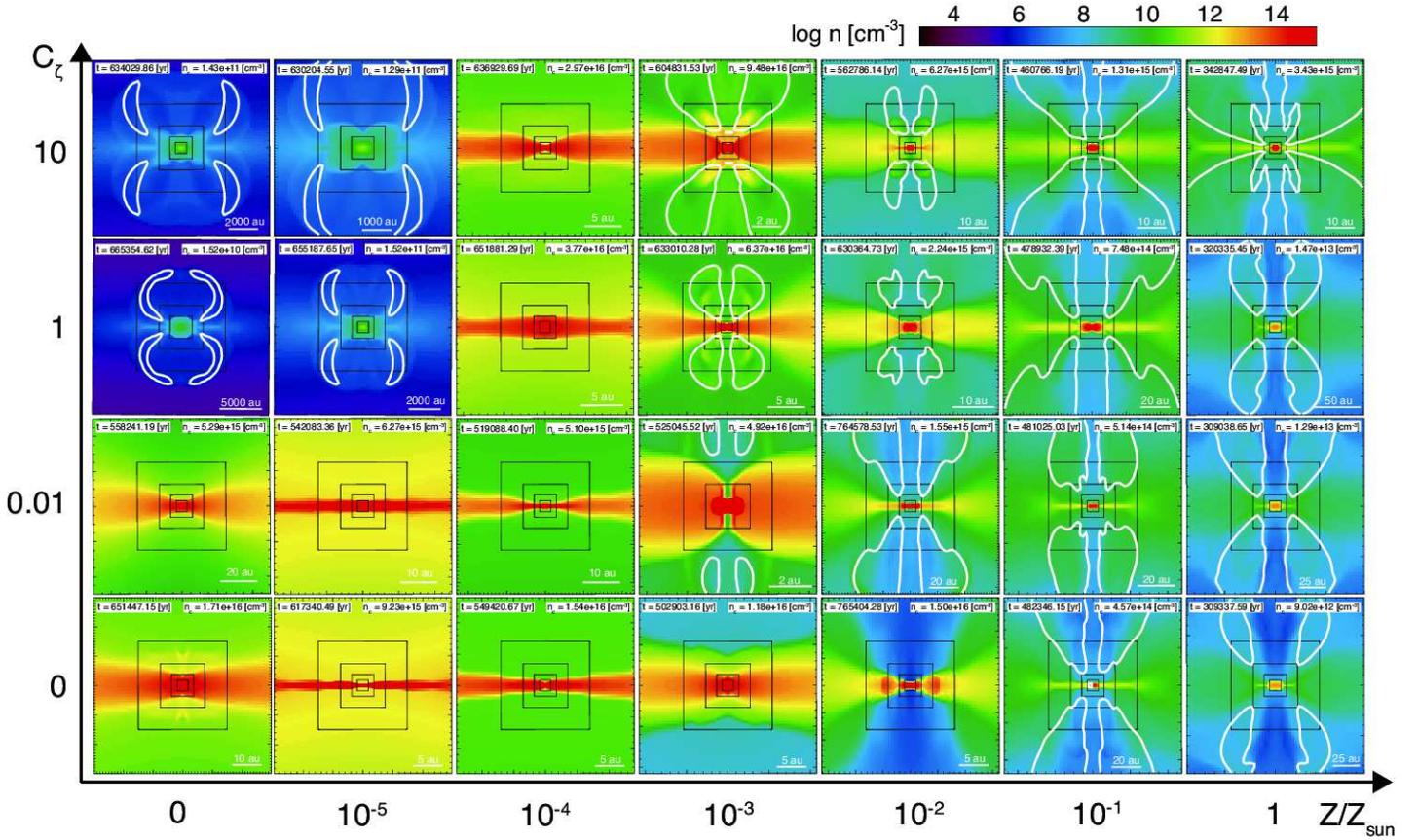}
\caption{
Plots of the density distribution of each model with models arranged based on their ionisation strengths $C_{\zeta}$ (vertical axis) and metallicities $Z/\zsun$ (horizontal axis). 
In each panel, the density distribution in the $y = 0$ plane is plotted, and the boundary between the inflow and outflow regions, within which the gas is outflowing from the central region, is plotted as a white contour in each panel.
The elapsed time and central number density are also given in the upper part of each panel, 
and the spatial scale is given in the bottom right corner of each panel.
}
\label{fig:outflow}
\end{figure*}
%\end{landscape}

\begin{figure*}
    \includegraphics[scale=0.50]{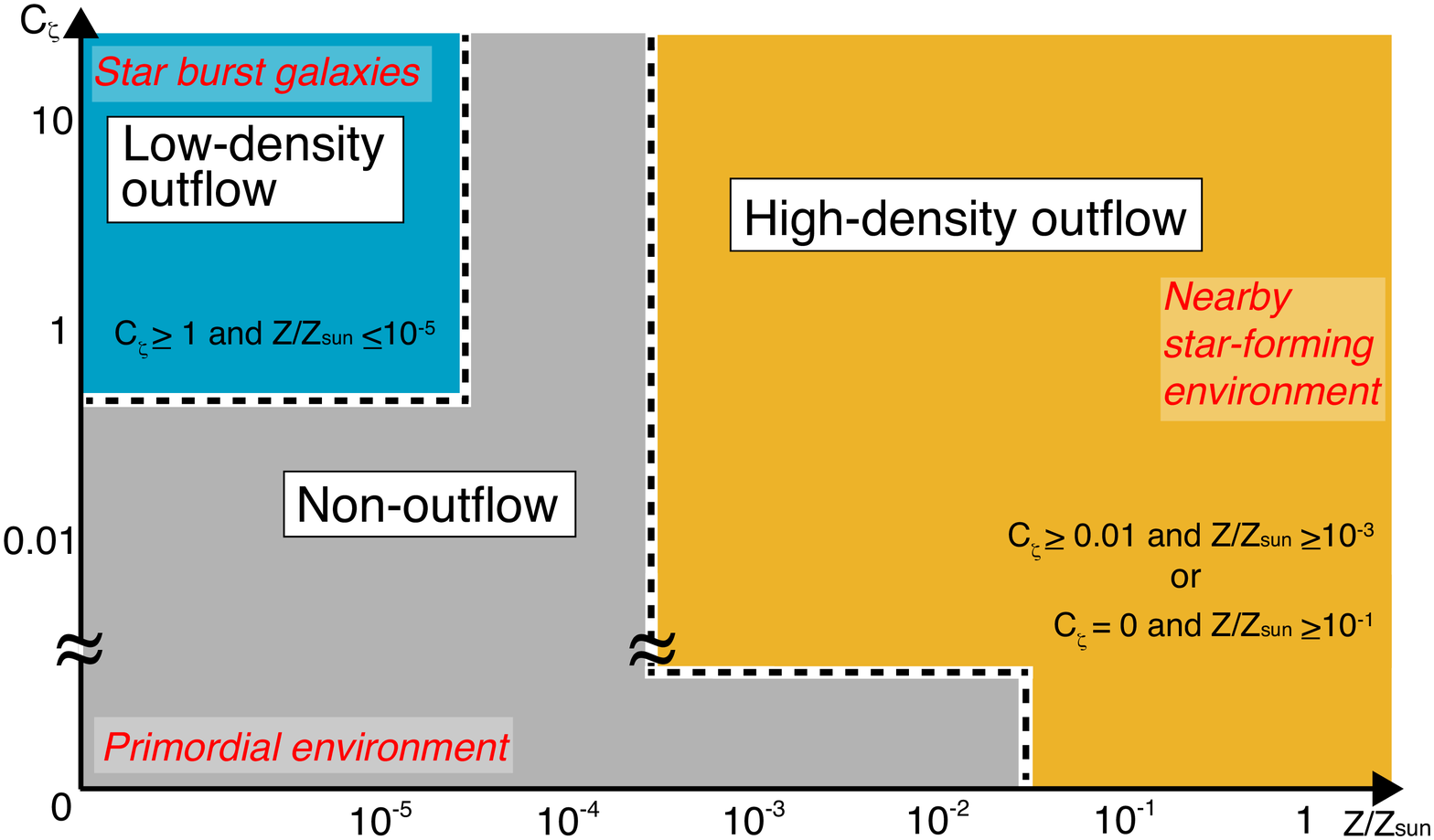}
\caption{
Simulation results. The vertical and horizontal axes are the same as in  Fig.~\ref{fig:outflow}. 
Models on blue background  show the outflow at a low density  ($n < 10^{10}\cm$; low-density outflow),  while models on orange background show it  at a  high  density ($n>10^{10}\cm$; high-density outflow).
Models on gray background  do not show the outflow by the end of the simulation. 
The parameter range for each flow (low- and high-density outflows) is described in each domain.  
}
\label{fig:outflow1}
\end{figure*}

\subsection{Outflow Driving in Parameter Space}
\label{sec:outflow}
As described in \S\ref{sec:model}, in some models, the outflow does not emerge until the end of the calculation. 
In the following, the models that show and do not show an outflow are referred to as `outflow' and `non-outflow' models, respectively. 
Fig.~\ref{fig:outflow} shows the outcomes for all the models listed in Table~\ref{table:1}.
Each panel of Fig.~\ref{fig:outflow} shows the density distribution in the $y=0$ plane when the outflow appears for the outflow models or when the central density reaches $n_{\rm c} \approx 10^{16}\cm$ for the non-outflow models.
The panels in the figure are arranged such that the vertical axis indicates the ionisation parameter $C_{\zeta}$ and the horizontal axis represents the metallicity $Z/\zsun$. 
We simply summarized the outcomes in Fig.~\ref{fig:outflow1}, in which the vertical and horizontal axes are the same as in Fig.~\ref{fig:outflow}.
The boundary between the outflow and non-outflow models is shown by black dashed lines in Fig.~\ref{fig:outflow1}.
The  models were thus classified into the following three types: 
 (i) models showing an outflow driven by a high-density region (orange in Fig.~\ref{fig:outflow1}), (ii) models showing an outflow driven by a low-density region (blue in Fig.~\ref{fig:outflow1}), and (iii) models not showing an outflow (gray in Fig.~\ref{fig:outflow1}). 
The outflow models categorised into types (i) and (ii) are thus hereafter referred to as `low-density outflow' and `high-density outflow' models, respectively. 
Note that since we only calculated the evolution of collapsing clouds before protostar formation (i.e. gas collapsing phase), the low- and high-density corresponds to the early and later epoch of the cloud evolution, respectively. 
In the low-density outflow models, the outflow emerges at a lower density and the outflow driving region, which roughly corresponds to the Jeans length at which the outflow emerges, is large. 
On the other hand, in the high-density outflow models, the flow emerges at a high density from a small disk-like structure. 
Here, we quantitatively define both outflow models.
In the low-density outflow models displayed on the blue background, the outflow emerges before the central density in the collapsing cloud reaches $n_{\rm c} =10^{10}\cm$.
In contrast, in the high-density outflow models displayed on the orange background, the outflow appears after the central density exceeds $n_{\rm c} > 10^{10}\cm$.
Figs.~\ref{fig:outflow} and \ref{fig:outflow1} indicate that the outflow tends to appear as the metallicity $Z/\zsun$ and ionisation strength $C_\zeta$ increase.

In the high-density outflow models (orange background), the outflow continues to be driven after the central number density reaches $n_{\rm c} \approx 10^{10}\cm$. 
Almost the same type of outflows can be seen in the present-day star formation simulations (i.e. $Z/\zsun \approx 1$), in which the magnetic field is strongly amplified due to the rotation immediately after the (first) adiabatic core formation \citep[][]{larson69,masunaga00} and the outflow is driven by magneto-centrifugal and magnetic pressure gradient forces \citep[][]{blandford82,uchida85,lynden-Bell03,banerjee06,machida08,bate14}. 
In some high-density outflow models, such as model I001Z3 ($Z/\zsun=10^{-3}$ and $C_\zeta=0.01$), the outflow seems to not be sufficiently evolved because the calculation was stopped when the number density reached $n_{\rm c} \approx 10^{16}\cm$. 
However, the outflow in such models would grow with time in a long-term calculation, as shown in previous works \citep{machida13,tomida17,matsushita17}.

For low-density outflow models (blue background), the outflow was initiated when the number density reached $n_{\rm c} \approx 10^{8}\cm$. 
The outflow in these models did not disappear until the end of the calculation. Although the cloud (central) density gradually increases, the outflow continues to appear in the region of $n\sim10^8\cm$ where an adiabatic core forms and sustains during the calculation (see below).
These flows tended to appear in models with lower metallicity and higher ionisation intensity.
In the cases with $C_{\zeta} = 1$ and $10$, which are relatively strong ionisation intensities,
the number of electrons increases, and  molecular hydrogen H$_2$ and deuterated molecular hydrogen HD are enhanced through the following reactions \citep[][]{galli98,stancil98,le bourlot99,flower00,flower02,flower03}: 
\begin{eqnarray}
& & e^- + {\rm H} \rightarrow {\rm H}^- + h\nu, \\
& & {\rm H}^- + {\rm H} \rightarrow {\rm H}_2 + e^-, \\
& & {\rm D}^+ + {\rm H}_2 \rightarrow {\rm HD} + {\rm H}^+, \\
& & {\rm D} + {\rm H}_2 \rightarrow {\rm HD} + {\rm H}.
\end{eqnarray}
The abundant H$_2$ and HD can reduce the gas temperature to below 100\,K in the range of $n_{\rm c} \lesssim 10^{8}\cm$ \citep[e.g.][]{uehara00,machida05b,nagakura09,hirano15,susa15}.
After the gas density of the collapsing cloud satisfies $n_{\rm } \gtrsim  10^{8}\cm$, molecular hydrogen further forms by the three-body reaction of hydrogen atoms \citep[e.g.][]{palla83,omukai05}, as
\begin{eqnarray}
{\rm H} + {\rm H } +{\rm H} \rightarrow {\rm H}_2 + {\rm H},\\
{\rm H} + {\rm H } +{\rm H}_2 \rightarrow {\rm H}_2 + {\rm H}_2.
\end{eqnarray}
Then, the gas temperature rapidly rises because of the heat of the reaction, and the gas pressure also increases (see Fig.~2 of Paper I).
Therefore, the gas collapse slows and the rotation timescale becomes shorter than the collapse timescale. 
As a result, a magneto-centrifugally or magnetic pressure driven outflow appears at this epoch. 
This is the mechanism by which the outflow is driven in low-metallicity and high-ionisation-intensity environments.
Although the outflow driving mechanism is almost qualitatively the same as in present-day star formation, the launching radius and density differ noticeably between low-density and high-density outflows. 
The former is because of the increase in temperature (or pressure) caused by the three-body reaction, whereas the latter is attributed to the inefficient dust cooling (see \S\ref{sec:color}).

%%%%%%%%%%
% Fig. 5
%%%%%%%%%%
\begin{figure*}
    \includegraphics[scale=0.6]{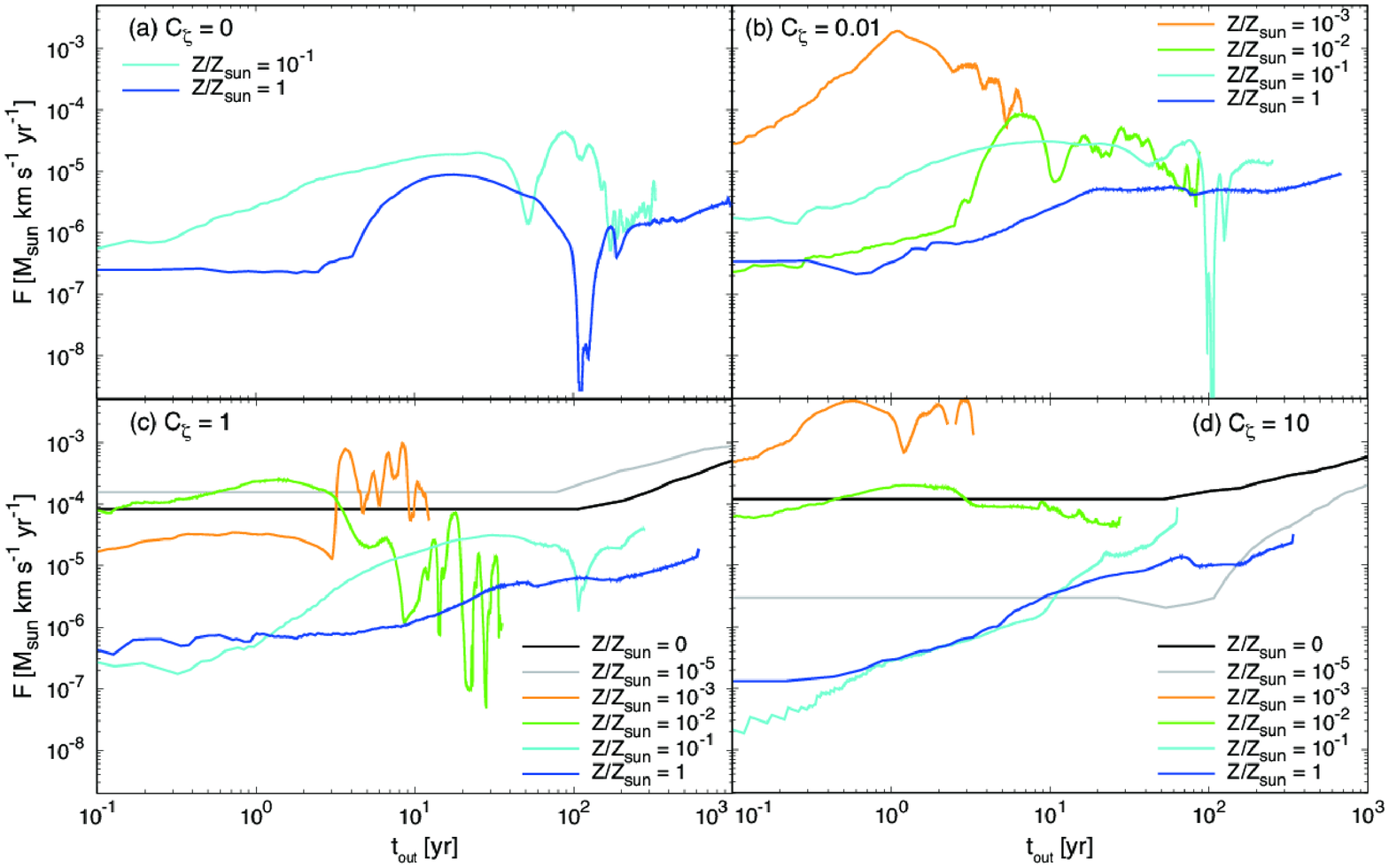}
\caption{
Outflow momentum flux plotted against the elapsed time after outflow emergence for the outflow models with ionisation strengths $C_\zeta$ of ({\it a}) $0$, ({\it b}) $0.01$, ({\it c}) $1$, and ({\it d}) $10$.
}
    \label{fig:mvt}
\end{figure*}

Fig.~\ref{fig:mvt} shows the outflow momentum flux, which is calculated as
\begin{equation}
F = \dfrac{\int^{v_{\rm r} >0} \rho \, v_{\rm r} \, \mathrm{d}V}{t_{\rm out}},
\end{equation}
where $v_{\rm r}$ ($>0$) is the outflow velocity and $t_{\rm out}$ is the elapsed time after the outflow emerges.
The outflow momentum flux has dimensions of force, and is useful to compare the power of the outflows of the different models.
In Fig.~\ref{fig:mvt}{\it c} and {\it d}, the black and grey lines correspond to the outflow momentum fluxes for the low-density outflow models (models I1ZP, I1Z5, I10ZP and I10Z5). 
These models have relatively large outflow momentum  fluxes.
A large outflow momentum flux indicates the emergence of a powerful outflow. 
The momentum fluxes for almost all of the outflow models gradually increase with time, which indicates that the outflow does not weaken during the calculation.

The outflow momentum fluxes in the observation of Class 0 objects are in the range of $10^{-8} \lesssim F/(\msun\,{\rm km}\,{\rm s}^{-1}) \lesssim 10^{-3}$ \citep{bontemps96}.
Thus, the momentum fluxes derived in simulations agree with those in observations of present-day star formation or nearby star-forming regions.
It was also confirmed that a large fraction of the infalling gas is ejected by the outflow, in which the ratio $M_{\rm out} / M_{\rm in}$ of the outflowing to infalling mass reaches  approximately 0.2--0.9 for each model. The total outflowing and infalling masses $M_{\rm out}$ and $M_{\rm in}$ are respectively estimated as
\begin{equation}
M_{\rm out} = \int^{v_{\rm r} >0} \rho  \, \mathrm{d}V,
\end{equation}
and
\begin{equation}
M_{\rm in} = \int^{v_{\rm r} <0 \ {\rm and} \ r<L_{\rm out}} \rho  \, \mathrm{d}V,
\end{equation}
where $L_{\rm out}$ is the length of the outflow. 
Thus, the present results indicate that if an outflow emerges, it significantly affects the star formation process in various environments.

\section{Discussion}
\label{sec:discussion}

%%%%%%%%%%
% Fig. 6
%%%%%%%%%%
\subsection{Magnetic Field Evolution  and Outflow Driving Condition}
\label{sec:color}
\begin{figure*}
    \includegraphics[scale=0.70]{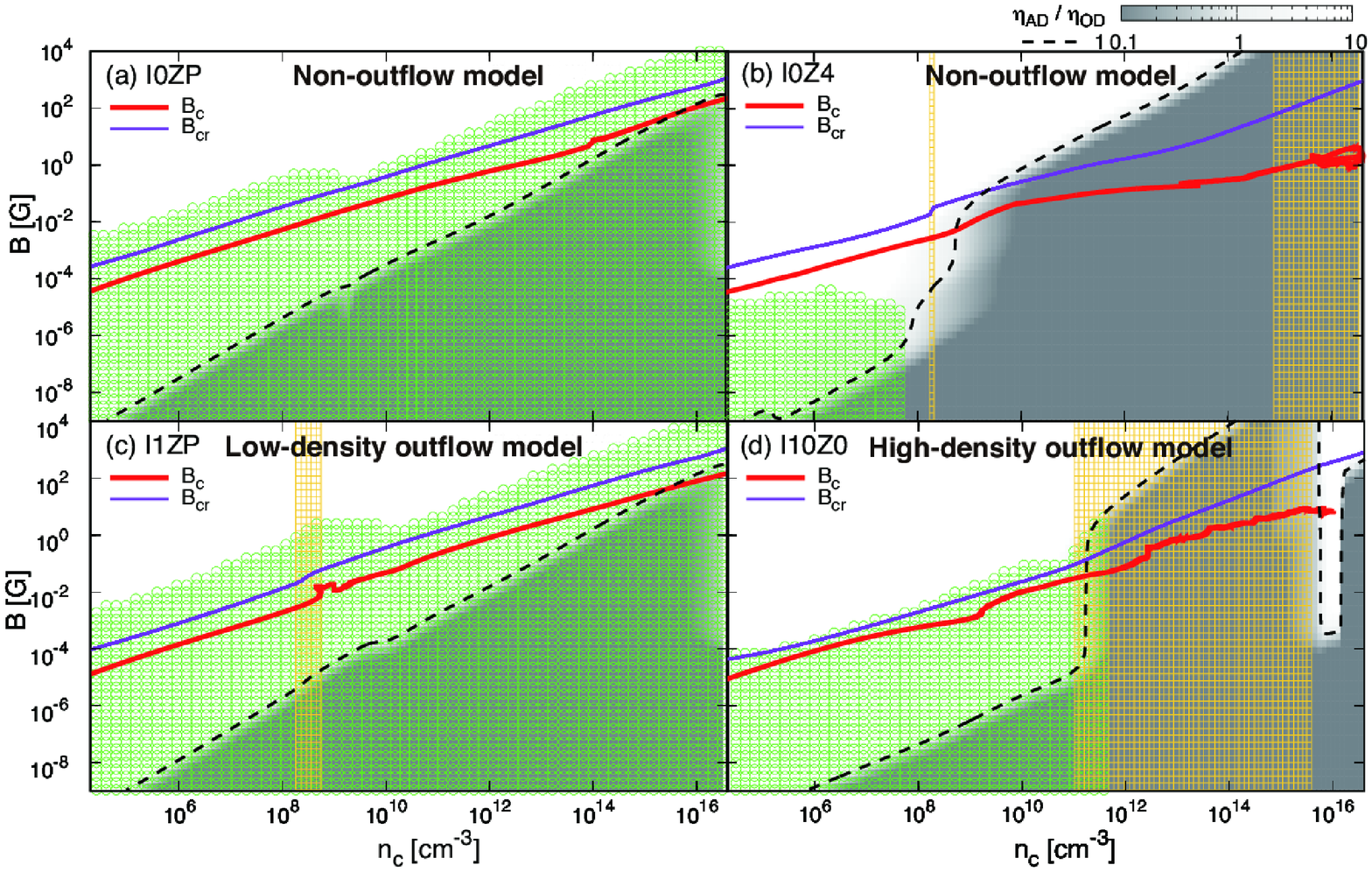}
\caption{
Evolution of the magnetic field strength $B$ at the cloud centre with respect to the central number density $n_{\rm c}$ for models (a) I0ZP, (b) I0Z4, (c) I1ZP, and (d) I10Z0. 
The critical magnetic field strength $B_{\rm cr}$ is plotted as a purple line (see Paper I).
The green region corresponds to the magnetically active region, in which the magnetic field is well coupled with neutrals, whereas the magnetic field is not coupled with neutrals in the other regions (grey and white regions).
The intensity of the grey colour in these regions represents the ratio of ambipolar diffusivity $\eta_{\rm AD}$ to Ohmic resistivity $\eta_{\rm OD}$.
The curve $\eta_{\rm AD}/\eta_{\rm OD}=1$ is plotted as a black broken line, above which ambipolar diffusion dominates Ohmic dissipation as the dissipation process of the magnetic field.
In the yellow hatched region, the adiabatic index $\gamma$ exceeds $4/3$ ($P\propto \rho^\gamma$), and a quasi-hydrostatic core can form in this region (see also Paper I).
}
    \label{fig:color}
\end{figure*}

Figs.~\ref{fig:color}{\it a}--{\it d} show the evolution of the central magnetic field for non-outflow models I0ZP and I1ZP, 
the low-density outflow model I1ZP, and the high-density outflow model I10Z0, respectively.
The present analysis revealed that an outflow appears in a collapsing cloud when the following three conditions are fulfilled: 
\begin{description}
\item[(a)] a stable adiabatic core forms before protostar formation,
\item[(b)] the stable core has a long lifetime, and
\item[(c)] the magnetic field is well coupled with the neutral gas around or inside  the stable core.
\end{description}
Conditions (a) and (b) are fulfilled when the effective polytropic index $\gamma$ exceeds $4/3$, which corresponds to the hatched yellow region in Fig.~\ref{fig:color}, and the collapsing cloud remains in the region of $\gamma>4/3$ for a long duration, where $\gamma$ is defined as $P\propto \rho^\gamma$ (see also Paper I). 
The gas thermal pressure (gradient force) can dominate the gravity and  temporarily stop (or slow) the cloud contraction with $\gamma > 4/3$. 
When there exists a wide density range satisfying $\gamma>4/3$, a quasi-hydrostatic core can form \citep[][]{larson69,masunaga00}. 
However, the mass accretion onto the quasi-static  core gradually increases the central density even in the region of $\gamma > 4/3$.
Again, the rapid collapse occurs when  the central density enters in the region of $\gamma < 4/3$.

When a long-lived quasi-static core appears and the magnetic field is well coupled with the neutral gas around the core (condition (c)), the rotation timescale becomes shorter than the collapse timescale.
This causes the magnetic field lines to become strongly twisted and amplified.
Finally, the outflow is driven by the rotating quasi-static (stable) core  by the magnetic effect. 
This process has been definitively established in the present-day star formation process \citep[e.g.][]{blandford82,tomisaka98,tomisaka00,tomisaka02,matsumoto04,machida08}.
In contrast, when the quasi-static core does not have sufficient time to amplify the magnetic field or  no quasi-static core forms in the collapsing cloud, no outflow appears. 
In addition, even when a long-lived core forms, no outflow appears unless the magnetic field is well coupled with the neutral gas (or unless condition (c) is fulfilled).

In Fig.~\ref{fig:color}, a stable hydrostatic core appears in the yellow hatched regions ($\gamma>4/3$), and the magnetic field is well coupled with the neutral gas in the green regions.
Thus, the outflow appears only in areas where the green region (magnetically active region) and the hatched yellow region (static core formation region) overlap, that is, where the magnetic field is well coupled with the neutral gas and is strongly amplified by the rotation of the quasi-static core.

The outflow does not appear before protostar formation in the primordial star formation case (Fig.~\ref{fig:color}{\it a}), which is consistent with \citet{machida08b}.
For this model, neither an adiabatic core nor an outflow appear because the condition $\gamma>4/3$ is not fulfilled under any conditions and no adiabatic core appears before protostar formation.
For model I0Z4 (Fig.~\ref{fig:color}{\it b}), although an adiabatic core appears, no outflow appears.
This is because, as shown in Fig.~\ref{fig:color}{\it b}, the hatched yellow region never overlaps with the green region, indicating that the magnetic field dissipates around or inside the core. 
On the other hand, there exists a region in which the green and yellow regions overlap in models I1ZP (Fig.~\ref{fig:color}{\it c}) and  I10Z0  (Fig.~\ref{fig:color}{\it d}), and thus outflows appear in both models.

\subsection{Dissipation of Magnetic Field and Angular Momentum Transfer}
\begin{figure*}
    \includegraphics[scale=0.50]{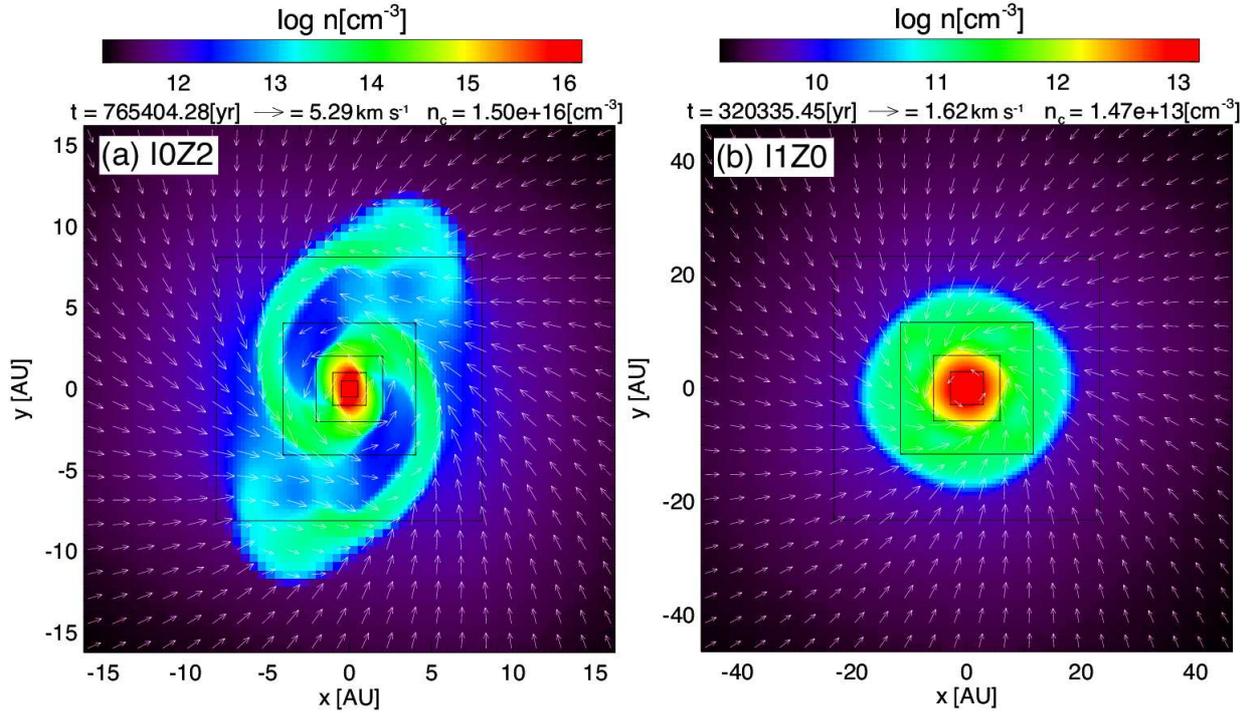}
\caption{
Density (colour) and velocity (arrows) distributions in the $z = 0$ plane for (a) the non-outflow model I0Z2 and (b) the outflow model I1Z0. 
The elapsed time $t$ and central density $n_c$ are given above each panel.
}
    \label{fig:GI}
\end{figure*}

Figs.~\ref{fig:GI}(a) and (b) show the structures of the adiabatic cores formed in the collapsing clouds  for the non-outflow model I0Z2 and the outflow model I1Z0, respectively.
In the present-day star formation process, the outflow is considered to transfer a large part of the angular momentum of the collapsing cloud into interstellar space \citep[e.g.][]{tomisaka98,tomisaka02,matsushita17}. 
As shown in Fig.~\ref{fig:GI}(b), the gas spirals into the central region in the outflow model because the angular momentum is effectively removed from the central region by magnetic effects such as outflow and magnetic braking.

In contrast, for the non-outflow model I0Z2 (Fig.~\ref{fig:GI}(a)), two spiral arms are visible. 
For this model, the angular momentum is not effectively transferred by magnetic effects because the magnetic field dissipates outside and inside the adiabatic core. 
Therefore, the adiabatic core, which is partly supported by the rotation, gradually increases in mass. 
Then, gravitational instability occurs and forms spiral arms \citep{toomre64}. 
For this model, fragmentation could not be confirmed.  
Fragmentation and binary or multiple star formation may occur in clouds embedded in similar environments.
Fig.~\ref{fig:GI} indicates that, in addition to the driving of the outflow, the structure around the cloud centre differs considerably among different star forming environments.

\subsection{Disk Formation: Magnetic Braking Catastrophe and Non-ideal MHD Effects}
This study focused on the gas collapsing phase prior to protostar formation, while we did not calculate the gas accretion phase following protostar formation. 
The circumstellar disk or Keplerian disk forms in the gas accretion phase following the gas collapsing phase. 
Thus, although we could not investigate the disk formation in this study, we briefly discuss the disk formation in the gas accretion phase.

In the present-day star formation process, some researchers pointed out that the angular momentum around the protostar is excessively transferred by magnetic braking  and circumstellar disk should not be formed in the early phase of the star formation (so-called magnetic braking catastrophe, \citealt{li14}).  
At that time, the numerical simulations imposing artificially large-sized sinks gave a misleading result, in which the circumstellar (or Keplerian) disks artificially disappear with  insufficient spatial resolutions \citep{machida14}. 
Then, since many Keplerian disks have been observed around Class 0 and I protostars by ALMA, now, the researchers have recognized it as no serious problem.
In addition, recent all state-of-art simulations including non-ideal MHD effects showed that the rotationally supported disk easily forms in a very early gas accretion phase \citep[see references in][]{tsukamoto16}, which well corresponds to the latest observations \citep[e.g.][]{yen17}.

On the other hand, the magnetic braking catastrophe may be serious in other star-forming environments. 
For example, as shown in Paper I, magnetic field does not significantly dissipate in the star forming clouds having a low metallicity and a strong ionization intensity, in which the magnetic field is well coupled with the neutral gas. 
Especially, when the star-forming cloud has a low metallicity and strong ionization intensity, non-ideal MHD effects are not effective.
In such environments, the magnetic braking may suppress the formation of  circumstellar disk. 
However, it should be noted that some researchers pointed out that disk formation is possible in strongly magnetized clouds without considering the dissipation of magnetic field when turbulence is present in the star forming cloud.
\citet{gray18} claimed that the rotationally supported disk forms when the rotation axis is not aligned with the magnetic field, in which the misalignment is caused by turbulence.
\citet{seifried12,seifried13} also showed that a turbulent environment can alleviate the magnetic braking catastrophe. 
These studies imply that the simple initial conditions of prestellar cloud, which has an uniform magnetic field and rigid rotation without turbulence, are not very appropriate to investigate the disk formation.
Thus, in addition to the dissipation of magnetic field, we need to consider turbulence in order to tackle the magnetic braking catastrophe.

Besides, the studies about the present-day star formation have shown that the outflow appeared in the gas collapsing phase continues to be driven in the gas accretion phase \citep[e.g.][]{machida13}.  
Both the outflow and magnetic braking mainly transfer the angular momentum in the gas accretion phase. 
Although the formation and evolution of the circumstellar disk is now beyond the scope of this study, we will focus on it in successive studies.

Finally, we comment on the Hall effect. 
There are three non-ideal MHD effects; Ohmic dissipation, ambipolar diffusion and Hall effect. 
The former two, which are considered in this study,  weaken the magnetic field  in the collapsing cloud, while the Hall term only changes the direction of magnetic fields. 
Thus, it was considered that the Hall effect is not very important in the star formation process.
However, recently, \citet{tsukamoto15b,tsukamoto17} showed that the Hall effect can control the size of the rotationally supported disk with the generation of toroidal fields in the collapsing cloud. 
Although the Hall coefficient strongly depends on the dust properties \citep{koga19}, it might affect the formation and evolution of the circumstellar disk in the gas accretion phase. 
Thus, we also have to carefully consider the Hall effect to investigate the disk formation in the gas accretion phase in future studies.

\subsection{Transition of Star Formation Mode}
Stars have been successively formed from the early universe to the present.
Theoretical studies have shown that the typical mass of stars formed from pristine gas (i.e. the first stars or primordial stars) is as high as approximately $10$--$1000\msun$ \citep[e.g.][]{hirano15}. 
Because of a lack of efficient coolant, the pristine gas has a high gas temperature, which leads to a large Jeans mass and the resultant formation of massive stars. 
In contrast, in present-day star formation, both dust and metal cooling play a role in determining the gas temperature and Jeans mass. 
Because the Jeans mass for prestellar clouds is as small as approximately $1 \msun$, the typical mass of present-day stars is $\lesssim 1\msun$. 
Thus, primordial and present-day stars have considerably different masses. 
Researchers have believed that the mass transition from approximately $100\msun$ (primordial environment) to approximately $1\msun$ (present-day environment) occurred at a certain epoch in the history of the universe. 
In past studies, it has been considered that the mass transition is characterised by the so-called critical metallicity $Z_{\rm{cri}}$, below which very massive stars preferentially form. 
Factors considered to contribute to the determination of $Z_{\rm cri}$ include dust \citep[e.g.][]{schneider02,schneider06,schneider12,omukai05,omukai10,dopcke13} and metal line cooling \citep[e.g.][]{Bromm01,Bromm03,Santoro06,Frebel07}. 
Realistic three-dimensional simulations have also been performed on this issue \citep[][]{jappsen07,Clark08,jappsen09a,jappsen09b,dopcke11,dopcke13,chiaki16}.
The results of these previous studies suggest that the critical metallicity lies in the range of $Z_{\rm{cri}} \approx 10^{-5}$--$10^{-3} \zsun$, above which cloud fragmentation occurs because of efficient cooling.

Before the transition of the star formation mode is discussed here, present-day star formation is addressed.
The gas temperature in the collapsing cloud remains at approximately $10$\,K before the dust cooling becomes optically thick at $n_{\rm c}\approx10^{11}\cm$.
After the gas cloud becomes opaque, the adiabatic core (or first core) forms. 
The central density of the adiabatic core gradually increases as the core mass increases as a result of mass accretion, and a second collapse occurs because of the dissociation of the molecular hydrogen at $n_c\approx10^{16}\cm$. 
Finally, a protostar appears after the dissociation finishes. 
Theoretical studies on present-day star formation have shown that a wide-opening outflow is driven by the first core and pulls a large fraction of the cloud mass back to interstellar space \citep{Matzner00,machida13}. 
Because the first core evolves into a rotationally supported disc (or circumstellar disc) after protostar formation, the outflow continues to be driven \citep{bate98,machida10}.
Thus, wide-angle outflows have a significant impact on present-day star formation.

Now, star formation in clouds with different metallicities is discussed. 
When the cloud metallicity is very low,\footnote
{
It was assumed that the dust abundance is proportional to the metallicity abundance (see \citet{susa15} and Paper I).
}
the first core cannot form because of the absence of dust grains. 
The first core clearly appears in clouds with a metallicity of $Z/\zsun \gtrsim10^{-4}$--$10^{-3}$ \citep{machida08b,susa15}. 
As described in \S\ref{sec:results}, this metallicity threshold also gives the condition of outflow launching.
It is not surprising that this condition roughly corresponds to the critical metallicity $Z_{\rm cri}/\zsun=10^{-5}$--$10^{-3}$ derived by the fragmentation condition, because both processes are
closely related to the formation of the first core. 
As a result, even from the perspective of the magnetic effects and the outflow, it may be concluded that the star formation mode changes at $Z/\zsun=10^{-4}$--$10^{-3}$ because the outflow expels a large fraction of the star forming gas into the interstellar space and significantly reduces the final stellar mass.

\section{Summary}
\label{sec:summary}
In the series of studies including the present work, the star formation process in different environments was investigated by considering the magnetic field and its dissipation, with the cloud metallicity $Z/\zsun$ and ionisation intensity $C_\zeta$ considered as parameters.
In Paper I, it was shown that the amplification and dissipation of the magnetic field strongly depends on the star formation environment. 
The magnetic field continues to be amplified without dissipation in star-forming clouds with an extremely low metallicity or strong ionisation intensity. 
However, the magnetic field dissipates both by Ohmic dissipation and ambipolar diffusion in high-metallicity and/or low-ionisation-intensity environments.

The present study focused on the driving of the outflow in collapsing clouds embedded in different star-forming environments. 
It was found that an outflow appears when a long-lived (first) adiabatic core forms in the magnetically active region where the magnetic field is well coupled with neutral gas. 
Roughly speaking, the transition of the outflow driving behaviour occurs between $Z/\zsun=10^{-4}$--$10^{-3}$, above which the outflow appears, whereas the transition metallicity depends somewhat on the ionisation intensity. 
The outflow tends to appear in clouds with a higher metallicity and stronger ionisation intensity.
A high metallicity allows the formation of a long-lived (first) adiabatic core, which drives the outflow.
A strong ionisation intensity suppresses the dissipation of the magnetic field and strengthens the coupling of the magnetic field and the neutral gas.

In addition, the results revealed a new type of outflow (low-density outflow), which is driven by the lower-density region (approximately $10^8\cm$) where the three-body reaction of hydrogen atoms heats the gas (see Fig.~1 of \citealt{susa15}) and slows the gas contraction, as in the formation process of the first adiabatic core. 
Then, the rotation timescale becomes shorter than the collapse timescale, and the outflow appears,  as shown in the upper left panels of Fig~\ref{fig:outflow1}.
The momentum flux in outflows of this type (low-density outflows) is comparable to that of high-density outflows.
Thus, the outflows seen in such an environment, which have a low metallicity and strong ionisation intensity, may impact star formation.

In this study, calculations were terminated when the central density reached approximately $10^{16}\cm$. 
Thus, the protostar itself could not be resolved, and the main accretion phase was not calculated. 
We need further long-term calculations to determine whether the outflows shown in this study sustain during the main accretion phase in future work.

\section*{Acknowledgements}
This study has benefited greatly from discussions with K. Doi. 
We also thank the reviewer for many useful comments on this paper. 
The present research used the computational resources of the HPCI system provided by (Cyber Sciencecenter, Tohoku University; Cybermedia Center, Osaka University, Earth Simulator, JAMSTEC) through the HPCI System Research Project (Project ID:hp160079, hp170047,hp180001, hp190035). 

The present study was supported by JSPS KAKENHI Grant Numbers
JP17K05387, JP17H02869, JP17H01101, and JP17H06360.
Simulations reported in this paper were also performed by 2017 Koubo Kadai on
Earth Simulator (NEC SX-ACE) at JAMSTEC.
This work was partly achieved through the use of supercomputer system SX-ACE at the Cybermedia Center, Osaka University.
Simulations were also performed by 2018 Koubo Kadai on Earth Simulator (NEC SX-ACE) at JAMSTEC. 

%%%%%%%%%%%%%%%%%%%% REFERENCES %%%%%%%%%%%%%%%%%%

%%%%%%%%%%%%%%%%% APPENDICES %%%%%%%%%%%%%%%%%%%%%
%%\appendix
%\section{Some extra material}

%If you want to present additional material which would interrupt the flow of the main paper,
%it can be placed in an Appendix which appears after the list of references.

%%%%%%%%%%%%%%%%%%%%%%%%%%%%%%%%%%%%%%%%%%%%%%%%%%

% Don't change these lines
%%\bsp	% typesetting comment
%%\label{lastpage}

\end{document}